# Tunable moiré bandgap in hBN-aligned bilayer graphene device with in-situ electrostatic gating

Hanbo Xiao[1]†, Han Gao[1]†, Min Li[1]†, Fanqiang Chen[2]†, Qiao Li[1], Yiwei Li[3], Meixiao Wang[1,4], Fangyuan Zhu[5], Lexian Yang[6], Feng Miao[2], Yulin Chen[1,7], Cheng Chen[1]*, Bin Cheng[8]*, Jianpeng Liu[1]*, Zhongkai Liu[1]*

[1]*School of Physical Science and Technology, ShanghaiTech Laboratory for Topological Physics, ShanghaiTech University, Shanghai 201210, China.*
[2]*Nanjing National Laboratory of Solid State Microstructures, School of Physics, Institute of Brain-Inspired Intelligence, Collaborative Innovation Center of Advanced Microstructures, Nanjing University, Nanjing 210093, P. R. China.*
[3]*Institute for Advanced Studies, Wuhan University, Wuhan, Hubei, 430072, P. R. China.*
[4]*Center for Transformative Science, ShanghaiTech University, Shanghai 201210, China.*
[5]*Shanghai Synchrotron Radiation Facility, Shanghai Advanced Research Institute, Chinese Academy of Sciences, Shanghai 201204, China.*
[6]*State Key Laboratory of Low Dimensional Quantum Physics, Department of Physics, Tsinghua University, Beijing, 100084, China.*
[7]*Department of Physics, Clarendon Laboratory, University of Oxford, Parks Road, Oxford OX1 3PU, UK.*
[8]*Institute of Interdisciplinary Physical Sciences School of Science, Nanjing University of Science and Technology, Nanjing 210094, P. R. China.*

*Corresponding author. Email: chencheng1@shanghaitech.edu.cn (C.C.), bincheng@njust.edu.cn (B.C.), liujp@shanghaitech.edu.cn (J.L.), liuzhk@shanghaitech.edu.cn (Z.L.)
†These authors contributed equally to this work.

**Over the years, great efforts have been devoted in introducing a sizable and tunable band gap in graphene for its potential application in next-generation electronic devices. The primary challenge in modulating this gap has been the absence of a direct method for observing changes of the band gap in momentum space. In this study, we employ advanced spatial- and angle-resolved photoemission spectroscopy technique to directly visualize the gap formation in bilayer graphene, modulated by both displacement fields and moiré potentials. The application of displacement field via in-situ electrostatic gating introduces a sizable and tunable electronic bandgap, proportional to the field strength up to 100 meV. Meanwhile, the moiré potential, induced by aligning the underlying hexagonal boron nitride substrate, extends the bandgap by ~ 20 meV. Theoretical**

**calculations, effectively capture the experimental observations. Our investigation provides a quantitative understanding of how these two mechanisms collaboratively modulate the band gap in bilayer graphene, offering valuable guidance for the design of graphene-based electronic devices.**

**Introduction**

The band gap in a semiconductor is one of the crucial factors that determines many electrical and optical properties of the system and key to the device performance, for instance, the on/off ratio of a field-effect transistor and the absorption wavelength of a photodetector. In recent years, there has been a substantial interest in the research of two-dimensional (2D) materials due to their versatile physical properties(*1–4*). Specifically, these materials exhibit bandgaps spanning from terahertz/mid-infrared to the visible and ultraviolet regions(*5–7*), making them promising candidates for various device applications.

Graphene stands out as a leading 2D material for electronic applications due to its unique properties, including unparalleled electrical and thermal conductivities, high electron mobility, and remarkable mechanical flexibility(*8, 9*). However, the absence of a bandgap in the intrinsic graphene system severely limits its potential applications(*10*). To create a sizable and controllable bandgap, great efforts have been devoted such as electrostatic gating(*5, 11–15*), atom intercalation(*16*), substrate doping(*17*), size confinement(*18*), rhombohedral stacking(*19*), and the introduction of moiré potentials(*1, 7, 20–31*). These bandgap tuning mechanisms could effectively modulate the gap size and tune the performance of graphene-based devices.

Understanding the interplay of these tuning methods requires a detailed investigation of their impact on graphene's physical properties, including changes in lattice structure, impurity levels, and electronic band structure. Although various characterization techniques (mostly indirect) including transport measurements(*2, 6, 12, 32–36*), scanning tunneling microscopy(*15*), and optical methods(*5, 37–40*) etc., have provided insights into the effects of these tuning methods, a comprehensive and direct observation of the momentum-space evolution of the band structure under multiple tuning knobs remains elusive. Such direct visualization is crucial for unraveling the underlying principles of the bandgap engineering in graphene, which could provide essential insights into the development of graphene-based electronic devices.

In this study, exploiting state-of-the-art spatial- and angle-resolved photoemission spectroscopy (NanoARPES) with in-situ electrostatic gating, we systematically investigate the electronic structure and the evolution of the bandgap in a Bernel-stacked bilayer graphene (BLG) device, in the presence of both moiré potential (through the moiré lattice formed by hexagonal boron nitride (hBN) substrate aligned with graphene layers) and displacement field (D-field, via electrostatic gating). By comparing the result with hBN-unaligned BLG devices, we find that while the band gaps in both BLG devices could be tuned continuously by the D-field up to 100 meV, the moiré potential could further enhance the band gap size by ~ 20 meV, which is consistent with our theoretical calculations. These findings provide a direct momentum-space visualization of the corroborative modulations of the gap size in bilayer graphene tuned by both D-field and moiré potential. Specifically, our results emphasize the significance of the moiré potential as an effective tuning mechanism for inducing substantial band gaps in graphene systems, and its potential in engineering the electronic structure of other applicable 2D devices.

**Results**

We conducted NanoARPES measurements to explore the electronic structure of both hBN-aligned and unaligned BLG devices. The experiment setup is illustrated in Fig. 1A (3D schematic) and 1B (schematic side view of the sample). The BLG device is grounded through the top Au contact, and a graphite/Au electrode is placed at the bottom of the hBN to serve as a back gate, which enables the electrostatic tuning of both the D-field and electron filling in BLG. The moiré potential in BLG is introduced by intentionally aligning the underlying hBN substrate to the graphene layers (with a twisted angle of 0.3º, see details in Methods). The small lattice mismatch between graphene and hBN enables the formation of a moiré lattice with a real-space periodicity ($L_{\text{moiré}}$) around 12 nm (Fig. 1C). Correspondingly, the moiré lattice generates mini-Brillouin zones (mini-BZs) around K point of BLG with a reciprocal vector ($G_{\text{moiré}}$) of 0.053 Å$^{-1}$ (Fig. 1D), which replicates the Dirac-shape band dispersion of BLG into wider momentum space (illustrated in Fig. 1E).

In Fig. 2, we present the NanoARPES result performed on hBN-aligned BLG device. The optical microscope image and the real-space intensity map of NanoARPES are presented in Fig.

2A, respectively, where the BLG/hBN regions can be precisely located. A representative constant energy contour taken at -0.6 eV below Fermi level ($E_F$) is illustrated in Fig. 2B(i) (complete electronic structure can be found in Fig. S1), where the main band structure of BLG and its moiré replicas are observed. This result is well reproduced by our simulation, considering both the effect of moiré potential and photoemission matrix element(*41*, *42*) on the BLG band structure (Fig. 2B(ii), details see section S1). The effect of moiré potential on the Dirac-shape main bands is more prominently seen in the band dispersion cuts (Fig. 2C for both ARPES and simulation), where the moiré replicas are observed on both sides, in sharp contrast with the hBN-unaligned situation (presented later in Fig. 3). The spacing between the main band and its moiré replica is measured to be 0.046 Å$^{-1}$, consistent with the twist angle of 0.3º. The charge neutral point (CNP) is found to locate slightly higher than $E_F$, thus only accessible with electrostatic gating.

The in-situ electrostatic gating with positive back gate voltage ($V_{bg}$) is then applied on the device, which effectively introduces more electrons into the system and enables the visualization of conduction band. Consecutive dispersion cuts taken with increasing $V_{bg}$ is presented in Fig. 2D, illustrating the evolution of the band structure. Besides the rigid shift of chemical potential, the presence of D-field brings in energy drop across the two layers of the BLG system, and a band gap at CNP is introduced consequently, whose size monotonically increases with $V_{bg}$, in line with previous reports(*14*). A representative dispersion cut (taken at $V_{bg}$ = 15V) is shown in Fig. 2E, where the size of the band gap is extracted to be around 100 meV (labeled by $\Delta_k$). Notably, the moiré replicas of the main bands are persistent in the presence of D-field, and no signature of additional in-gap state is observed (see further analysis in Fig. S4).

To better understand the effect of moiré potential on BLG system, we conducted comparative NanoARPES measurement on BLG device unaligned with hBN substrate, and the result is presented in Fig. 3. The twist angle between BLG and hBN is examined to be around 15° (see Fig. S3 for details) in this device, corresponding to a moiré periodicity of 9.5 nm in real space. In this situation, the effect of moiré lattice on the band structure is considered minimum, due to the large moiré vector in momentum space ($G$ = 0.66 Å$^{-1}$), thus weak Fourier component of the moiré potential, as compared to the dispersive Dirac-shape band of graphene in the vicinity of

K point. Indeed, the constant energy contour taken at -0.6 eV below $E_F$ (Fig. 3B for both ARPES and simulation, complete electronic structure is illustrated in Fig. S2) shows characteristic feature of intrinsic BLG with two concentric rings, deriving from the two Dirac-shape bands of BLG with an energy difference around 400 meV (Fig. 3C for both ARPES and simulation). No signature of moiré modulation on the electronic structure, as compared with hBN aligned case in Fig.2, is evidenced here. Comparatively, we conduct the in-situ electrostatic gating on this device (Fig. 3D), and it shows similar behavior including both the shift of chemical potential and the opening of the band gap (Fig. 3E).

To quantitatively distinguish the effects of moiré potential and D-field in BLG system, we extract the sizes of the band gap in each ARPES spectrum and plot them as a function of the D-field $D = \epsilon V_{bg}/2d$ (Fig. 4A, here $\epsilon$ is the dielectric constant and $d$ is the thickness of hBN, see section S2 for details). The evolution of the band gap in hBN-aligned and unaligned BLG devices shows a similar trend, increasing monotonically with the enhancement of D-field. Notably, the size of the band gap in hBN-aligned BLG system is found to be larger than the unaligned case by $\Delta E \sim 20$ meV with the same D-field, indicating the additional enhancement of band gap from the moiré potential. To capture these experimental findings in BLG/hBN system, theoretical calculation is performed based on the tight-binding model (see Method for details). As shown in Fig. 4B, the simulated band structure of hBN-aligned BLG suggests that the moiré potential can sufficiently introduce a band gap $\sim 20$ meV at the CNP, due to the $C_{2z}$ symmetry breaking. This is notably different from the monolayer situation, where the hBN substrate can only open the band gap by 2 meV(*43*). In addition, the band gap in hBN-aligned BLG system could be further enhanced by D-field, as it increases the difference of the interlayer potential in BLG system. By introducing the D-field in the model (see Method for details), the simulated spectrum (Fig. 4C) well captures the experimental result, in which the band gap increases with the D-field while the main band and its moiré replicas remain stable. The theoretical results are summarized in Fig. 4D (the unaligned data is reproduced from previous reports(*4*, *5*)), and in general captures our experimental findings.

**Discussion**

Our measurement of the moiré band gap in the hBN-aligned BLG is consistent with previous observations using other techniques such as STM (~ 9 meV)(*44*), optical (38 meV)(*45*) and photocurrent spectroscopy (~ 14 meV)(*25*). The slight difference, i.e., 20 meV vs. 14 or 9 meV may results from the different twist angles and/or local strains(*46–48*) in the different devices. It is worth noting that the difference in bandgap size between the hBN-aligned and the unaligned samples remains constant with increasing D-field, which implies that the overall bandgap can be interpreted as the arithmetic sum of the contributions from the moiré potential and the D-field induced gaps.

The overall bandgap in our bilayer graphene (BLG) system thus can be conceptualized as the cumulative effect of two distinct mechanisms: interlayer potential drop and intralayer $C_{2z}$ symmetry breaking. The D-field applied across the BLG establishes differential potential energies across its two layers, while alignment with hBN substrate breaks the $C_{2z}$ rotational symmetry. The interaction between graphene and the aligned hBN is well characterized by the moiré potential, which gives rise to secondary moiré replica bands. Together, these mechanisms cooperatively influence the bandgap in the hBN-aligned BLG system, resulting in the gate tunable modulation of its electronic properties.

The observed combined-effect of the moiré potential and D-field may help to understand the unique transport properties at the hBN/BLG interface, for example, the ratcheting mechanism for carrier injection in hBN/BLG based moiré synaptic transistor devices(*49*). Meanwhile, it is proposed that the moiré potential may generate "localized" carriers point at the aligned hBN/BLG interface(*50*). Our measurement on the hBN-aligned BLG provides momentum-space spectroscopic evidence about the moiré bands whose dispersions remain intact across the large range of the D-field, which helps reveal the nature of the interface states.

In summary, using state-of-the-art NanoARPES system equipped with in-situ electrostatic gating, we systematically investigated the evolution of electronic structure of BLG systems both aligned and unaligned with hBN substrate. The moiré potential modulated electronic structure of BLG is directly revealed, with the presence of an energy gap at CNP of ~ 20 meV, resulting from the $C_{2z}$ symmetry breaking induced by moiré potential. The presence of D-field can further increase this band gap with a rate of $d\Delta_k/dD$ ~ 0.11 eV per V/nm, due to enhancement of interlayer potential energy difference. Our measurement provides direct spectroscopic evidence of bandgap engineering under multiple tuning knobs in BLG/hBN systems and addresses the importance of moiré potential in reshaping the electronic structure of the system. The synergistic integration of multiple tuning methods in BLG systems offers a versatile approach to altering the electronic structure of BLG. This modification can be further refined through the application of external fields, strain, or substrate interactions, thereby enhancing the design and functionality of graphene-based electronic devices.

**Materials and Methods**

**Device Fabrication:** We fabricated the hBN-aligned bilayer graphene by a modified polymer-based pick-up technique. Graphene and hBN flakes were exfoliated on separate SiO2/Si substrates, with the thickness of these flakes identified with optical contrast and atomic force microscope (Bruker MultiMode 8). Thickness of hBN is 16.2 nm for unaligned sample and 19.0 nm for aligned one. A polycarbonate–poly(dimethylsiloxane) (PC/PDMS) stamp was used to pick the bilayer graphene (BLG), hBN and graphite flakes at 80°C sequentially. Then the stack was released on the SiO2/Si substrate at 140°C. The polycarbonate film was dissolved by chloroform. The angle between BLG and hBN was measured by AFM and further checked by ARPES. The contact electrodes (Cr 5 nm/Pd 15 nm/Au 40 nm) to BLG and bottom graphite gates were deposited by standard electron-beam evaporation after patterned by electron beam lithography.

**NanoARPES measurement with in-situ electrostatic gating:** The nano-ARPES measurements were performed at the BL07U endstation of Shanghai Synchrotron Radiation Facility (SSRF). The base pressure is lower than $1\times10^{-10}$ mbar. The sample was annealed at 300 °C for 3 hours prior to the measurement to desorb absorbates. The beam was focused using a Fresnel zone plate (FZP) and a spatial resolution of ~ 400 nm was achieved. All the experiments were performed at T = 20 K with a photon energy of 92.6 eV and right-handed circular polarized light with a fixed incidence angle of 60°. The data were collected by a hemispherical Scienta DA30 electron analyzer. The instrumental energy and momentum resolution under the measurement conditions was ~10 meV and 0.01 Å, respectively.

**Atomistic tight-Binding Model:** A realistic atomic tight-binding model is used to study the electronic structure of bilayer graphene/hBN. In the atomistic tight-binding model, an empirical Slater Koster-type parameter $t(d)$ was used to describe the hopping between the $p_z$ atomic

orbitals at different sites(51)

$$t(\boldsymbol{d}) = V_\sigma^0 e^{-\frac{(r-d_c)}{\delta_0}} \left(\frac{\boldsymbol{d}\cdot\hat{z}}{d}\right)^2 + V_\pi^0 e^{-\frac{(r-a_0)}{\delta_0}} \left[1 - \left(\frac{\boldsymbol{d}\cdot\hat{z}}{d}\right)^2\right] \quad (1)$$

where $V_\sigma^0 = 0.48$ eV, $V_\pi^0 = -2.7$ eV, and $d_c$ is 3.35 Å representing the interlayer distance of graphene. $\delta_0 = 0.184a$, where a is the lattice constant of graphene and $a_0 = a/\sqrt{3}$. $d$ is the displacement vector between two sites. We assume the on-site potential $V_C = 0$ for carbon atom, and $V_B = 3.34$ eV, $V_N = -1.40$ eV for boron and nitride atoms, respectively. The lattice structure is rationalized to the relative lattice period $a_{hBN}/a \approx 1.018$ to 56/55 with assumed perfect alignment $\theta = 0°$.

**Spectral Function Calculations:** To simulate the ARPES intensity, the unfolded spectral functions are calculated for the bilayer system along $k$ path in the atomic BZ near the K valley, and the unfolded spectral function is expressed as(52)

$$A(\omega, \boldsymbol{k}) = -\frac{1}{\pi} \text{Im}\left[\sum_{l\alpha}\sum_n \frac{|\langle l\alpha \boldsymbol{k} | n\boldsymbol{k}\rangle|^2}{\omega - E_{n\boldsymbol{k}} + i\delta}\right] \quad (2)$$

where $l$ refers to the layer index and $\alpha$ represents the sublattice index. $|n\boldsymbol{k}\rangle$ is the Bloch wave function at the $\boldsymbol{k}$ point of the $n$-th moiré energy band, and $|l\alpha\boldsymbol{k}\rangle$ denotes the Bloch wave function of the $l$-th monolayer graphene and $\alpha$ sublattice, without the moiré potential effects. The $n$-th moiré Bloch function at wavevector $\boldsymbol{k}$ could be expressed as

$$|n\boldsymbol{k}\rangle = \sum_{il\alpha} C_{il\alpha,n}(\boldsymbol{k})|il\alpha\boldsymbol{k}\rangle \quad (3)$$

where $i$ stands for the atomic lattice vector index within a moiré primitive cell. The Bloch-like state $|il\alpha\boldsymbol{k}\rangle$ at wavevector $\boldsymbol{k}$ could be expressed in the basis of the $p_z$-orbital-like Wannier functions as

$$|il\alpha\boldsymbol{k}\rangle = \frac{1}{\sqrt{N_M}} \sum_R e^{i\boldsymbol{k}\cdot\boldsymbol{R}} |il\alpha\boldsymbol{R}\rangle \quad (4)$$

where $\boldsymbol{R}$ is the moiré lattice vector. Then, we obtain:

$$\langle l\alpha\boldsymbol{k} | n\boldsymbol{k}\rangle = \sum_{\boldsymbol{R}'i'l'\alpha'} \langle l\alpha\boldsymbol{k} | i'l'\alpha'\boldsymbol{R}\rangle\langle i'l'\alpha'\boldsymbol{R} | n\boldsymbol{k}\rangle = \frac{1}{\sqrt{N_0}} \sum_i e^{-i\boldsymbol{k}\cdot\boldsymbol{a}_i} C_{il\alpha,n}(\boldsymbol{k}) \quad (5)$$

in which $\boldsymbol{a}_i$ refers to the $i$-th atomic lattice vector within a moiré primitive cell for the $l$-th layer, and $N_0$ is the number of atomic unit cells within a moiré primitive cell.

**Acknowledgments**

Z. K. Liu acknowledges the support from the National Natural Science Foundation of China (92365204, 12274298) and the National Key R&D program of China (Grant No. 2022YFA1604400/03). B. Cheng acknowledges the support from the National Natural Science Foundation of China (12322407, 12074176).

S. H. Zhang was supported by the National Natural Science Foundation of China (12304217) and the Fundamental Research Funds for the Central Universities from China

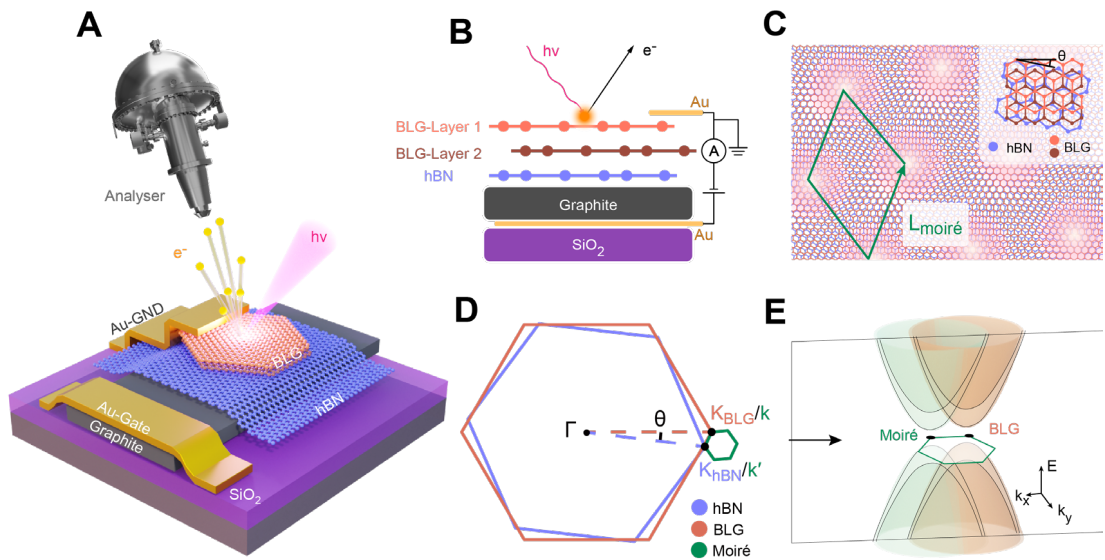

**Fig. 1. NanoARPES on BLG/hBN heterostructure. (A)** Schematics of the experimental setup. **(B)** Side-view of the sample and gating geometries. **(C)** Real-space illustration of BLG lattice with almost aligned hBN substrate with a twist angle of $\theta$. The generated moiré unit cell is indicated by the green rhombus. **(D)** Illustration of the Brillouin zones (BZs) of hBN (blue), BLG (orange). The mini-BZ generated from the moiré lattice is labelled in green. $K_{BLG}$/ $K_{hBN}$ and $k/k'$ label the high symmetry points in the extended BZ and the mini-BZ, respectively. **(E)** Schematic showing the replication of Dirac-shape main bands by the moiré potential in the mini-BZ.

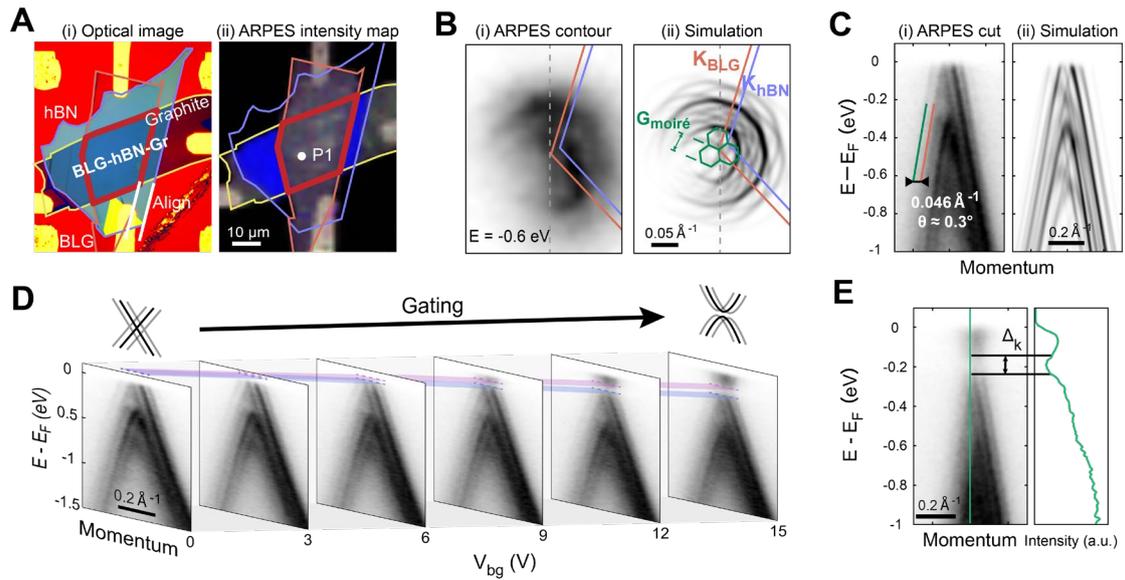

**Fig. 2. Electronic structure of aligned BLG/hBN and its evolution with in-situ electrostatic gating.**
**(A)** Real space information of the aligned BLG/hBN sample. (i) Optical image and (ii) NanoARPES real space intensity map. Data presented in (B-E) was taken at P1. **(B)** Constant energy contours near the $K_{BLG}$ at $E = E_F$ - 0.6 eV for (i) ARPES and (ii) simulation(*41*). BZ of BLG, hBN and mini-BZ are labelled. **(C)** Band dispersion across $K_{BLG}$ along the direction indicated by the grey dashed line in (B), for (i) ARPES and (ii) simulation. The momentum distance between BLG band and its replica (orange and green line) is measured to be 0.046 Å$^{-1}$, corresponding to a twist angle of 0.3° between hBN and BLG. **(D)** Evolution of the band dispersion with consecutive back gate voltages. The color stripes indicate the evolution of the CNP and opening of the bandgap. **(E)** Band dispersion cut taken at $V_{bg}$ = 15 V and the energy distributions curve taken at K point (green line), where the band gap $\Delta_k$ is extracted.

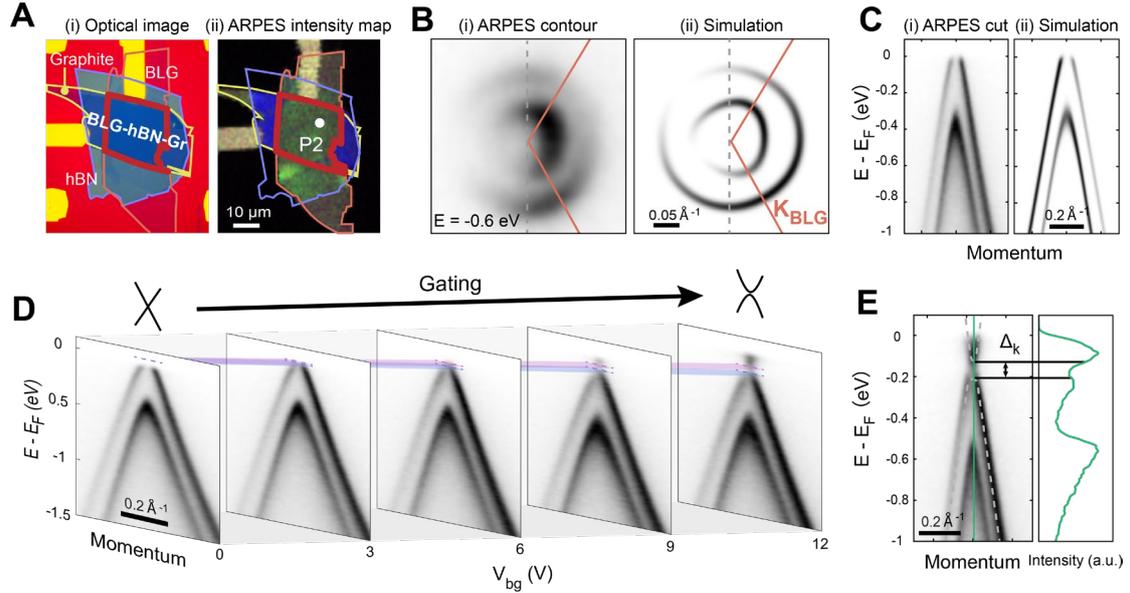

**Fig. 3. Electronic structure of unaligned BLG/hBN and its evolution with in-situ electrostatic gating.** **(A)** Real space information of the unaligned BLG/hBN sample. (i) Optical image and (ii) NanoARPES real space intensity map. ARPES measured at point P2. The boundary of each segment is indicated with the same color in Fig.1. **(B)** Constant energy contours near the $K_{BLG}$ at $E = E_F - 0.6$ eV of (i) ARPES and (ii) simulation(*41*). BZ of BLG is labelled. **(C)** Band dispersion across $K_{BLG}$ along the direction labelled by the grey dashed line in (b), by (i) ARPES and (ii) simulation. **(D)** Evolution of the band dispersion across $K_{BLG}$ along the direction labelled by the grey dashed line in (C) at different back gate voltage. The color stripes indicate the evolution of the CNP and opening of the bandgap. **(E)** Band dispersion across $K_{BLG}$ along the direction labelled by the grey dashed line in (B) with $V_{bg} = 12$ V (left panel) and the plot of the energy distributions curve at K point (right panel). $\Delta_k$ labels the band gap between the valence and conduction band.

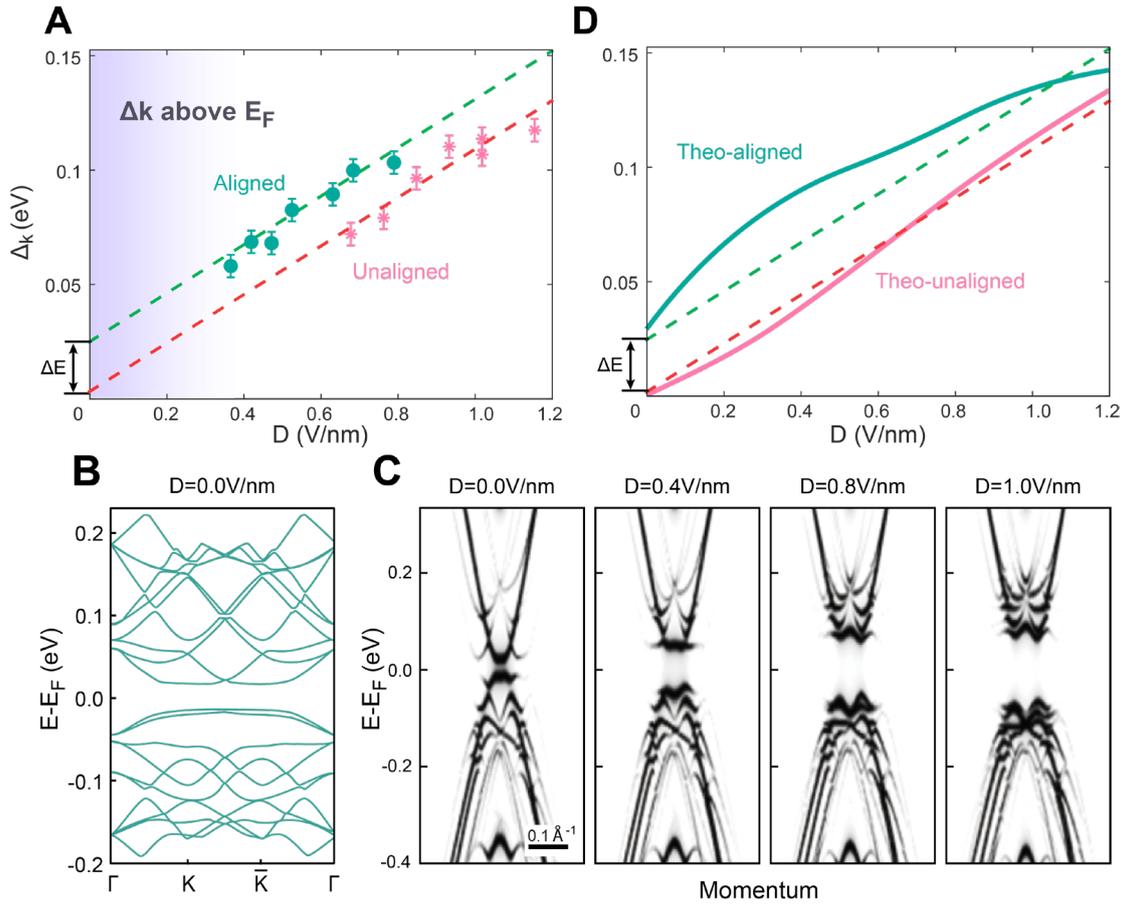

**Fig. 4. Calculated spectral function and bandgap evolution as a function of D field. (A)** Plot of the bandgap $\Delta_k$ extracted from NanoARPES spectra for aligned (green) and unaligned (pink) BLG/hBN as a function of D field. The dashed lines indicate the trend of the bandgap evolution, and the energy difference is labeled by $\Delta E \sim 20$ meV. **(B)** Calculated band structures of aligned BLG/hBN system with $\theta = 0°$ based on tight-binding model. The high symmetry points are within the moiré mini-BZ. **(C)** Simulated band dispersion cut along the same direction as Fig. 2c with D field for $D = 0, 0.4, 0.8$ and $1.0$ V/nm, respectively. **(D)** Plot of the bandgap $\Delta_k$ extracted from theoretical calculation for aligned (green) and unaligned (pink) BLG/hBN as a function of D field. The dashed lines mark the trend of the bandgap evolution taken from experiments as in (A).

**Supplementary Materials**

**This PDF file includes:**

Sections S1 to S2

Figs. S1 to S4